# Point defects and dopants of boron arsenide from first-principles calculations: donor compensation and doping asymmetry

S. Chae, K. Mengle, J. T. Heron, and E. Kioupakis[a]

*Department of Materials Science and Engineering, University of Michigan, Ann Arbor, MI 48109, USA*

We apply hybrid density functional theory calculations to identify the formation energies and thermodynamic charge transition levels of native point defects, common impurities, and shallow dopants in BAs. We find that boron-related defects such as $V_B$, $B_{As}$, $B_i$-$V_B$ complexes, and antisite pairs are the dominant intrinsic defects. Native BAs is expected to exhibit p-type conduction due to the acceptor-type characteristics of $V_B$ and $B_{As}$. Among the common impurities we explored, we found that C substitutional defects and H interstitials have relatively low formation energies and are likely to contribute free holes. Interstitial hydrogen is surprisingly also found to be stable in the neutral charge state. $Be_B$, $Si_{As}$ and $Ge_{As}$ are predicted to be excellent shallow acceptors with low ionization energy (< 0.03 eV) and negligible compensation by other point defects considered here. On the other hand, donors such as $Se_{As}$, $Te_{As}$ $Si_B$, and $Ge_B$ have a relatively large ionization energy (~0.15 eV) and are likely to be passivated by native defects such as $B_{As}$ and $V_B$, as well as $C_{As}$, $H_i$, and $H_B$. The hole and electron doping asymmetry originates from the heavy effective mass of the conduction band due to its boron orbital character, as well as from boron-related intrinsic defects that compensate donors.

[a] Author to whom correspondence should be addressed. Electronic mail: kioup@umich.edu

Boron arsenide (BAs), a III-V zincblende semiconductor, has received attention recently due to the theoretical prediction[1] and subsequent synthesis and experimental validation[2-4] of its unusually high thermal conductivity. The observed room-temperature thermal conductivity (1300 W m$^{-1}$ K$^{-1}$) surpasses all other bulk materials except diamond.[2] However, BAs has several potential advantages compared to diamond in terms of cost-effectiveness and compatibility with existing III-V semiconductor technology. Specifically, BAs may be useful as an active electronic component since it has a similar electronic structure to Si but has a wider band gap.[5] Moreover, BAs, as a member of the group III-V arsenide series (BAs-AlAs-GaAs-InAs) or boron series (BP-BAs-BN), can be applicable for alloying conventional III-V semiconductors.[6] Therefore, BAs is a promising material for applications in microelectronics for which efficient heat dissipation is crucial for the performance of devices.

In spite of its attractive thermal properties, the semiconducting properties of BAs have been relatively unexplored. The primary reason is the difficulty in fabricating reasonably large and pure single-crystal samples. However, millimeter-size crystals have recently been synthesized with chemical vapor transport[3] and enable characterization of its fundamental electronic properties.

Besides the intrinsic bulk properties of a semiconductor, native point defects and dopants are important for determining the characteristics of devices. For instance, donor and acceptor incorporation is an essential step in semiconductor technology to achieve the desired type and level of electrical conductivity. In addition, the perturbations of the crystal potential caused by defects decrease the electrical and thermal conductivity. First-principles calculations are a powerful tool to understand point-defect properties since experimental studies to identify and characterize defects at the atomic scale are challenging. Although the effects of intrinsic defects on the thermal conductivity of BAs have been previously explored,[7-8] there are no reports investigating the role of defects and dopants on its electrical properties.

In this work, we study the thermodynamic properties of point defects in BAs such as native defects, shallow dopants, and common impurity elements, using first-principles calculations. We identified the dominant native and common impurity defects, and we predict that BAs is inherently p-type. Our results show that extrinsic p-type conduction is easier to achieve in BAs than n-type conduction owing to the smaller acceptor ionization energy and negligible acceptor compensation.

We performed first-principles calculations based on hybrid density function theory (DFT) using the projector augmented wave (PAW) method and the Heyd-Scuseria-Ernzerhof (HSE06)[9] functional with 25% mixing as implemented in the Vienna Ab initio Simulation Package (VASP)[10-12]. The employed PAW pseudopotentials[13,14] include the B $2s^22p^1$ and As $4s^24p^3$ electrons in the valence with a cutoff energy of 400 eV. All bulk and defect structures were relaxed using the quasi-Newton algorithm with a maximal force criterion of 0.01 eV/Å. Point-defect calculations were performed using a cubic 64-atom supercell with a cell size of 9.54 Å. Spin polarization was considered for supercells with odd numbers of electrons. We performed supercell size convergence test by comparing to a 216-atom cell and found that the formation energy and transition energy of a Si atom that substitutes on the B site deviates by less than 0.1 eV after finite-cell-size energy corrections. We used a 2×2×2 Γ-centered mesh of k-points to sample the first Brillouin zone. However, for the ionization energy of shallow donors and acceptors, the total energies of their neutral and charged states were calculated with a refined 5×5×5 k-points mesh using the 2×2×2 k-mesh relaxed atomic structure to include states near the conduction band minimum [approximately near (0.8, 0, 0) in units of $2\pi/a$, where $a$ is the lattice constant]. The formation energy of a point defect D in charge state of q is determined by[15]:

$$E^f(D^q) = E_{tot}(D^q) - E_{tot}(BAs) - \sum n_i(E_i + \mu_i) + q(E_F + E_v) + E_{corr}(D^q), \tag{1}$$

where $E_{tot}(D^q)$ is the total energy of a supercell containing the defect, $E_{tot}(BAs)$ is the total energy of the pristine bulk supercell, $n_i$ is the number of defect atoms added to (or removed from) the supercell, and $\mu_i$ and $E_i$ are the chemical potential and energy per i atom in its elemental phase. For charged defects, the formation energy depends on the Fermi level ($E_F$), which is referenced to the valence-band maximum, $E_v$. We included an additional correction of 0.07 eV arising from the spin-orbit splitting of the topmost valence bands. The correction energy for the unphysical electrostatic interaction between periodic charged-defects images $E_{corr}(D^q)$ is calculated with the SXDEFECTALIGN code.[16] The static dielectric constant of the host material is $\varepsilon_0 = 9.6$, as determined with density functional perturbation theory[17] and the Quantum ESPRESSO code[18]. The chemical potentials, $\mu_i$, are determined by the growth conditions. The equilibrium condition for the formation of BAs is $\mu_B + \mu_{As} = \Delta H_f(BAs)$, where $\Delta H_f(BAs) = -0.238$ eV is our calculated formation enthalpy of BAs. The chemical potentials for As rich / B poor conditions are bounded by the formation of metallic As ($\mu_{As} = 0$ eV and $\mu_B = \Delta H_f(BAs)$) while As poor / B rich conditions are determined by the formation of elemental B ($\mu_{As} = \Delta H_f(BAs)$ and $\mu_B = 0$ eV). Other secondary phases we considered for the impurity formation energies are $BH_3$, h-BN, $C_2B_{13}$, $As_2O_5$, AsSi, $As_2Ge$, $As_2Se_3$, and $As_2Mg_3$, while the reference elemental phases are the bulk phases of B, As, C (graphite), Si, Ge, Se, Te, Mg and Be, and the $H_2$, $O_2$ and $N_2$ molecules. The calculated formation enthalpies of the secondary phases are listed in Table S1.

The calculated properties of bulk BAs are summarized in Table I. Our results show that the HSE hybrid functional with 25% mixing predicts the lattice parameter and enthalpy of formation of BAs in excellent agreement with experimental values. Our calculated band structure of BAs is shown in Fig. 1 (b). The direct and indirect band gaps are found to be 4.13 eV and 1.90 eV, respectively. For comparison, the experimental gap was estimated to be 1.46 eV from photocurrent measurements.[19] However, this result is inconclusive since it was measured from a discontinuous layer of BAs including boron gaps and oxygen contaminants.[19] In comparison with other theoretical results, LDA calculations predict the band gap to be 1.25 eV,[5] and recent GW calculations predict a value of 2.05 eV.[20] Our gap result is therefore in good agreement with the value obtained with the GW method.

We first examine intrinsic defects since they are unavoidable during growth and can dramatically affect the electrical and thermal properties. Intrinsic defects we considered include B vacancies ($V_B$), As vacancies ($V_{As}$), Schottky defects ($V_B$-$V_{As}$), B antisites ($B_{As}$), As antisites ($As_B$), antisite pairs ($B_{As}$-$As_B$), B interstitials ($B_i$), As interstitials ($As_i$), B Frenkel defects ($V_B$-$B_i$), and As Frenkel defects ($V_{As}$-$As_i$). Their atomic structures are shown in Fig. S1 and their formation energies are plotted as a function of the Fermi energy in Fig. 2 for both As-rich and B-rich conditions. Under both growth conditions, boron-related defects such as $B_{As}$, $V_B$, and $B_{As}$-$As_B$ are predicted to dominate in BAs owing to their lower relative formation energies. $B_{As}$ is a shallow acceptor with an ionization energy of 0.05 eV. It has the lowest formation energy for Fermi energies near CBM, indicating that it is an important donor-passivating defect. However, for Fermi energies near the VBM, $B_{As}$ more likely forms a pair with $As_B$ in the neutral charge state. $V_B$ is also an important native defect in BAs with a relatively low formation energy for Fermi energies near the CBM. It is stable in various charge states depending on the Fermi level, but it is a deep acceptor with the 0/-1 transition level at $E_F$ = 0.66 eV. On the other hand, $As_B$ has a high formation energy compared to the $B_{As}$-$As_B$ pair for most Fermi energies, indicating that it is more likely to be stabilized within antisite defect pairs. The concentrations of other defects (interstitials, As vacancies, Schottky defects) are expected to be negligible under any conditions because of their high formation energy.

We next investigate common impurity defects related to H, O, N, and C atoms, since these elements are common in the environment in the form of water, atmospheric air, or organic contaminants and are often inadvertently incorporated into a material during growth and post-processing. Our results for their formation energies at the two extreme growth conditions are shown in Fig. 3 and their atomic structures in Fig. S2, while their lowest formation energy over the entire range of chemical potentials is plotted in Fig. S3. Among the various types of impurity defects, we predict that substitutional C defects, both $C_{As}$ and $C_B$, form easily, possibly owing to the valence of C being the average of B and As. We find that $C_B$ acts like a deep donor with ionization energy of 0.47 eV, whereas $C_{As}$ is a shallow acceptor with a 0.09 eV ionization energy. However, in intrinsic BAs, C prefers to occupy the As site rather than the B site and pins the Fermi level from 0.15 eV to 0.40 eV above the VBM depending on the growth conditions. $H_i$ and $H_B$ are also likely to form in BAs due to the small size of H. $H_i$ has a relatively low formation energy throughout the entire Fermi-level range and its charge state varies from +1 to -1. In most semiconductors and insulators $H_i$ acts as a "negative-U" center that is only stable in the +1 and -1 charge states, and its charge-transition level serves as a universal band-alignment criterion.[21] We surprisingly find that the neutral charge state is also stable in BAs. This unusual result, also observed for interstitial hydrogen in diamond and boron nitride,[22] may be attributed to the small lattice constant of BAs compared to other III-V semiconductors. In addition to $C_{As}$ and $H_i$, $H_B$ in the -2 charge state has a low formation energy for Fermi energies near the CBM, showing that it is another important charge-compensating defect in n-doped BAs. The concentrations of other defects associated with these four common impurities, such as $H_{As}$, $C_i$ as well as O- and N-related defects, are negligibly small because of their high formation energies.

We subsequently investigate the possibility of n-type doping of BAs with Se, Te, Si, and Ge dopants. Se and Te are group-16 elements with ionic radii close to that of $As^{+5}$, and are thus expected to preferentially occupy the As site. On the other hand, since the ionic radii of the group-14 elements (Si and Ge) are comparable to both $B^{+3}$ and $As^{+5}$, we investigated Si and Ge atoms that substitute both the B-site as donors and the As-site as acceptors for comparison. Our results (Fig. 4 and Fig. S4) show that the donor ionization energy (i.e. the 0 to +1 charge transition level) is 0.16 eV for $Se_{As}$, 0.13 eV for $Te_{As}$, 0.14 eV for $Si_B$ and 0.17 eV for $Ge_B$. We therefore predict that, regardless of the type of donor atom, shallow donors have relatively large ionization energies in BAs. This is in contrast to Si and GaAs, both of which have much smaller donor ionization energies (~0.05 eV for As, P and Sb donors in Si[23] and ~0.006 eV for Se, Si, and Ge donors in GaAs[24]). We attribute the large ionization energy of BAs to its heavy electron effective mass that, according to the Bohr model of donor ionization, leads to large ionization energy and poor donor activation. We also note that although Se and Te preferentially substitute the As site, donating electrons, Si and Ge are more likely to occupy the As site as acceptors rather than serving as donors on the B site. Thus, we conclude that Si and Ge dopants preferentially act as acceptors in BAs. Overall, however, the formation energies of donors are higher than the formation energies of compensating negatively charged intrinsic defects such as $B_{As}$ and $V_B$, as well as negatively charged $C_{As}$, $H_i$, and $H_B$. Therefore, donors are highly likely to be passivated by boron-related intrinsic defects and potential unintentional impurities in BAs.

We also investigated p-doping of BAs by Be and Mg acceptors on the B-site, as well as Si and Ge acceptors on the As-site. Although we found that $Mg_B$ is energetically unfavorable due to the large ionic-size difference between Mg and B, we predict that $Be_B$, $Si_{As}$ and $Ge_{As}$ are excellent shallow acceptors with low ionization energy (< 0.03 eV) and sufficiently low formation energy (Fig. 4 and Table II). Their maximum formation energies occur at $E_F$ = VBM, which are 0.51 eV for $Be_B$, 0.91 eV for $Si_{As}$, and 1.35 eV for $Ge_{As}$ under chemical-potential conditions that minimize their formation energy (Fig. S5). Considering the higher formation energies of passivating donor-type intrinsic defects and common impurities (such as $As_B$, $C_B$ and $H_i$) at this Fermi-energy region, shallow acceptors are not likely to be compensated in BAs. We also explored any possible self-compensation from the dopant element themselves when incorporated into undesired sites. Fig. S6 shows the formation energies of acceptor species incorporated into interstitial sites of BAs. Interstitial dopants act as donors but they have much higher formation energies compared to substitutional defects, and thus do not incorporate in appreciable concentrations. In addition, Fermi pinning owing to Si or Ge dopants incorporated onto both the B and As sites occurs only for a specific range of chemical potentials. Under As-rich conditions, the formation energies of $Si_B$ and $Ge_B$ in the +1 charge state are slightly lower than $Si_{As}$ and $Ge_{As}$ for Fermi energies near the VBM, thus Be is the best acceptor candidate under these conditions. On the other hand, in As-poor conditions, Si and Ge are stable in the -1 charge state on the As site for the entire range of Fermi energies. Therefore, Be, Si, and Ge can all p-type dope BAs without charge compensation.

In conclusion, we investigated the thermodynamic properties of common defects, impurities, and dopants in BAs with first-principles calculations. Our results show that the unique physical properties of boron result in the doping asymmetry of BAs. The large electron effective mass of BAs that originates from the small radius of the B orbitals leads to relatively large ionization energies of shallow donors. Moreover, donor atoms have higher formation energy than acceptors and are compensated by negatively charged boron-related intrinsic defects and common impurities in BAs such as $B_{As}$, $V_B$, $C_{As}$, $H_i$, and $H_B$. On the other hand, excellent p-type dopability of BAs is predicted using Be, Si, and Ge elements with ionization energies less than 0.03 eV. Therefore, our results uncover the fundamental challenges and opportunities of doping BAs for applications in semiconductor devices.

TABLE I. Experimental values and HSE06 calculated values for the indirect band gap, lattice parameter, and formation enthalpy of BAs.

|  | Experimental | Calculated (this work) |
| --- | --- | --- |
| a (Å) | 4.78[2] | 4.77 |
| $\Delta H_f$ (eV/f.u.) | -0.408[25] | -0.238 |
| $E_g$ (eV) | 1.46[19] | 1.90 |

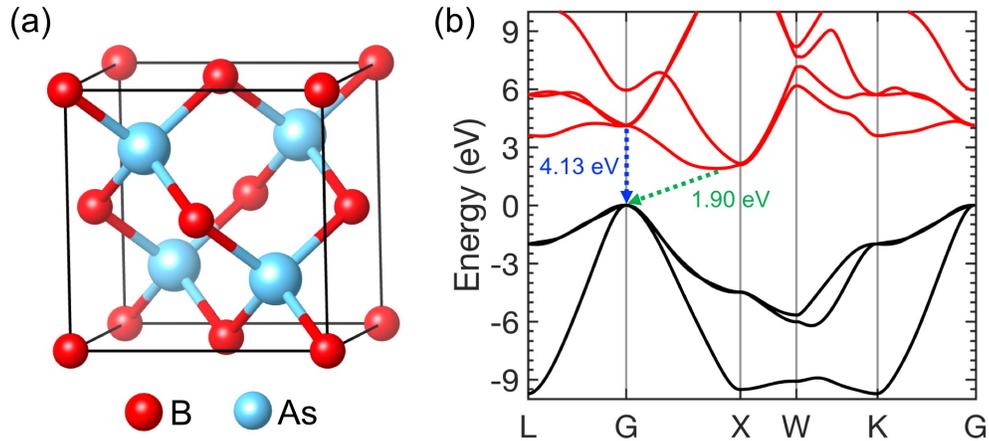

FIG. 1. (a) Zincblende crystal structure of BAs. (b) Calculated band structure of BAs using the HSE hybrid functional. The calculated indirect gap is 1.90 eV and the minimum direct gap at Γ is 4.13 eV.

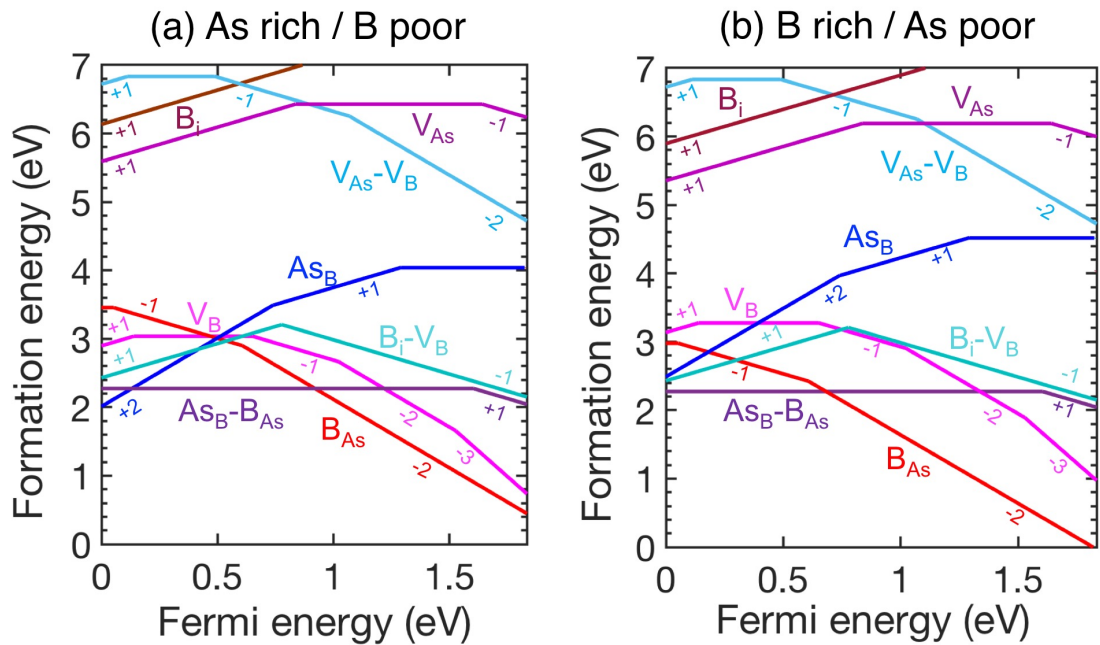

FIG. 2. Formation energy of various intrinsic point defects as a function of the Fermi level in the limit of (a) As-rich/B-poor and (b) B-rich/As-poor growth conditions.

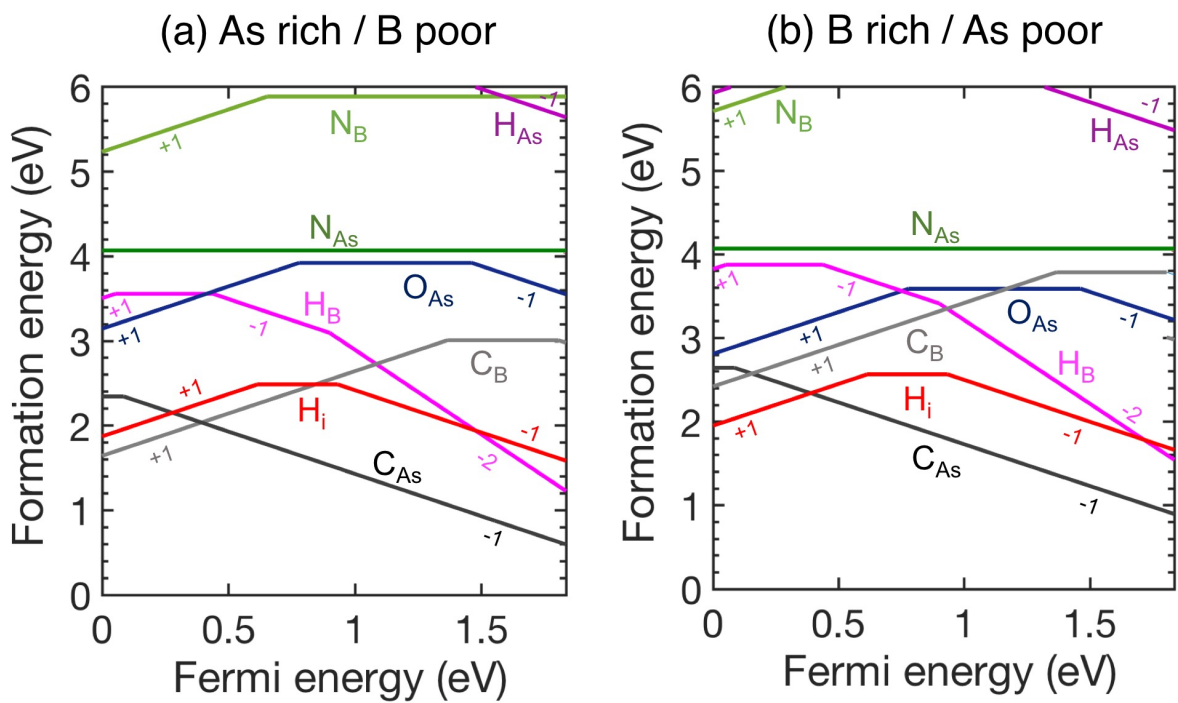

FIG. 3. Formation energy of common impurity defects as a function of the Fermi level in the limit of (a) As-rich/B-poor and (b) B-rich/As-poor growth conditions.

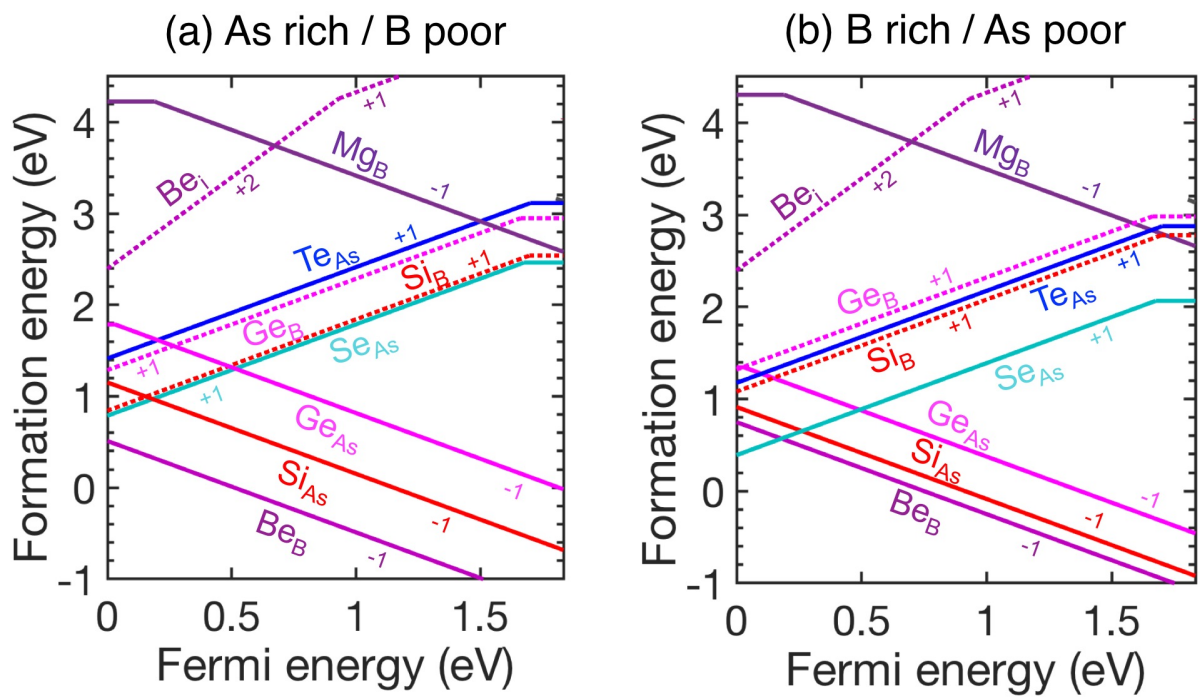

FIG. 4. Formation energy of donor and acceptor impurities as a function of the Fermi level in the limit of (a) As-rich/B-poor and (b) B-rich/As-poor growth conditions.

TABLE II. Shallow donor and acceptor ionization energies of BAs.

| Donor | Ionization Energy (eV) | Acceptor | Ionization Energy (eV) |
|---|---|---|---|
| $Se_{As}$ | 0.16 | $Be_B$ | < 0.03 |
| $Te_{As}$ | 0.13 | $Si_{As}$ | < 0.03 |
| $Si_B$ | 0.14 | $Ge_{As}$ | 0.03 |
| $Ge_B$ | 0.17 | $Mg_B$ | 0.19 |

## SUPPLEMENTARY MATERIAL

See supplementary material for the formation enthalpies of impurity phases, atomic structure of defects, the lowest formation energy of common impurities, extrinsic dopants, and interstitial dopants.

## ACKNOWLEDGMENTS

This work was supported by the Designing Materials to Revolutionize and Engineer our Future (DMREF) Program under Award No. 1534221, funded by the National Science Foundation. K.A.M. acknowledges the support from the National Science Foundation Graduate Research Fellowship Program through Grant No. DGE 1256260. This research used resources of the National Energy Research Scientific Computing Center, a DOE Office of Science User Facility supported by the Office of Science of the U.S. Department of Energy under Contract No. DE-AC02-05CH11231.

## REFERENCES

[1] L. Lindsay, D.A. Broido, and T.L. Reinecke, Phys. Rev. Lett. **111**, 025901 (2013).

[2] J.S. Kang, M. Li, H. Wu, H. Nguyen, and Y. Hu, Science **361**, 575 (2018).

[3] F. Tian, B. Song, X. Chen, N.K. Ravichandran, Y. Lv, K. Chen, S. Sullivan, J. Kim, Y. Zhou, T.H. Liu, M. Goni, Z. Ding, J. Sun, G.A.G.U. Gamage, H. Sun, H. Ziyaee, S. Huyan, L. Deng, J. Zhou, A.J. Schmidt, S. Chen, C.W. Chu, P.Y. Huang, D. Broido, L. Shi, G. Chen, and Z. Ren, Science **361**, 582

(2018).

[4] S. Li, Q. Zheng, Y. Lv, X. Liu, X. Wang, P.Y. Huang, D.G. Cahill, and B. Lv, Science **361**, 579 (2018).

[5] R.M. Wentzcovitch and M.L. Cohen, J. Phys. C Solid State Phys. **19**, 6791 (1986).

[6] G.L.W. Hart and A. Zunger, Phys. Rev. B **62**, 13522 (2000).

[7] Q. Zheng, C.A. Polanco, M. Du, R. Lucas, M. Chi, J. Yan, and B.C. Sales, Phys. Rev. Lett. **121**, 105901 (2018).

[8] N.H. Protik, J. Carrete, N.A. Katcho, N. Mingo, and D. Broido, Phys. Rev. B **94**, 045207 (2016).

[9] J. Heyd, G.E. Scuseria, and M. Ernzerhof, J. Chem. Phys. **118**, 8207 (2003).

[10] G. Kresse and J. Hafner, Phys. Rev. B **47**, 558 (1993).

[11] G. Kresse and J. Furthmüller, Comput. Mater. Sci. **6**, 15 (1996).

[12] G. Kresse and J. Furthmüller, Phys. Rev. B **54**, 11169 (1996).

[13] P.E. Blöchl, Phys. Rev. B **50**, 17953 (1994).

[14] D. Joubert, Phys. Rev. B **59**, 1758 (1999).

[15] C. Freysoldt, B. Grabowski, T. Hickel, J. Neugebauer, G. Kresse, A. Janotti, and C.G. Van De Walle, Rev. Mod. Phys. **86**, 253 (2014).

[16] C. Freysoldt, J. Neugebauer, and C.G. Van De Walle, Phys. Rev. Lett. **102**, 016402 (2009).

[17] S. Baroni, S. De Gironcoli, A. Dal Corso, S. Scuola, I. Superiore, I. Istituto, F. Materia, I.- Trieste, and P. Giannozzi, Rev. Mod. Phys. **73**, 515 (2001).

[18] P. Giannozzi, S. Baroni, N. Bonini, M. Calandra, R. Car, C. Cavazzoni, D. Ceresoli, G.L. Chiarotti, M. Cococcioni, I. Dabo, A. Dal Corso, S. De Gironcoli, U. Gerstmann, C. Gougoussis, A. Kokalj, M. Lazzeri, L. Martin-samos, N. Marzari, F. Mauri, R. Mazzarello, S. Paolini, A. Pasquarello, L. Paulatto, and C. Sbraccia, J. Phys. Condens. Matter **21**, 395502 (2009).

[19] S. Wang, S.F. Swingle, H. Ye, F.R.F. Fan, A.H. Cowley, and A.J. Bard, J. Am. Chem. Soc. **134**,

11056 (2012).

[20] K. Bushick, K. Mengle, N. Sanders, and E. Kioupakis, unpublished.

[21] C. G. Van der Walle and J. Neugebauer, Nature **423**, 626 (2003).

[22] J.L. Lyons and C.G. Van De Walle, J. Phys. Condens. Matter **28**, 06LT01 (2016).

[23] W. Kohn, Solid State Phys. **5**, 257 (1957).

[24] A.G. Milnes, *Deep Impurities in Semiconductors* (New York,NY: Wiley, 1973).

[25] H. Dumont and Y. Monteil, J. Cryst. Growth **290**, 410 (2006).

Supplementary Material

# Point defects and dopants of boron arsenide from first-principles calculations: donor compensation and doping asymmetry

S. Chae, K. Mengle, J. T. Heron, and E. Kioupakis[1]

*Department of Materials Science and Engineering, University of Michigan, Ann Arbor, MI 48109,*

*USA*

TABLE S1. HSE06 calculated values for the formation enthalpies of secondary phases considered for the determination of chemical potential constraints.

|  | $\Delta H_f$ (eV/f.u.) |
|---|---|
| $BH_3$ | -2.861 eV |
| h-BN | -0.408 eV |
| $C_2B_{13}$ | -1.077 eV |
| $As_2O_5$ | -4.336 eV |
| AsSi | -0.668 eV |
| $As_2Ge$ | -0.204 eV |
| $As_2Se_3$ | -0.668 eV |
| $As_2Mg_3$ | -0.376 eV |

[1] Author to whom correspondence should be addressed. Electronic mail: kioup@umich.edu

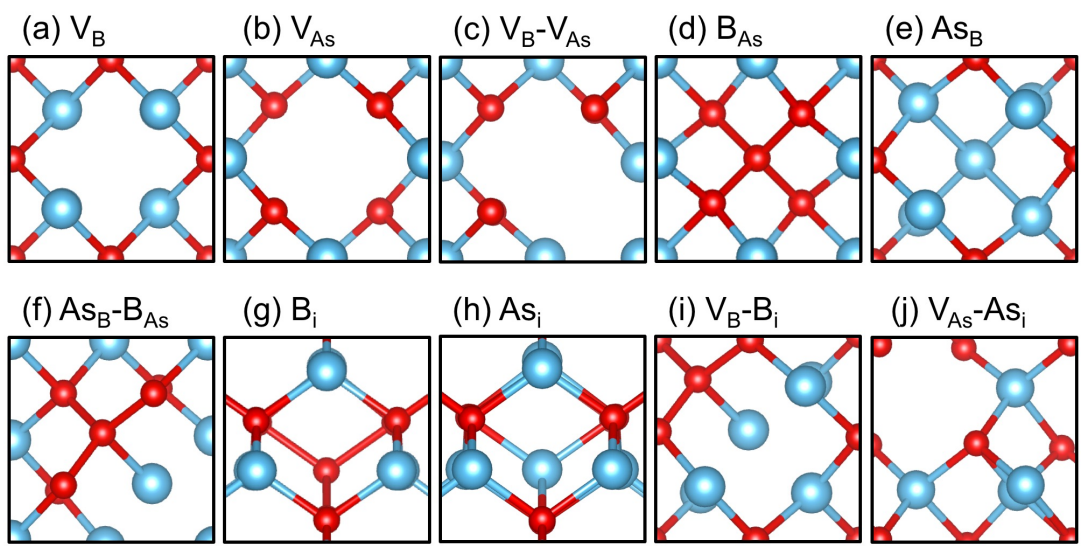

FIG. S1. Atomic structure of intrinsic defects in BAs after structure optimization

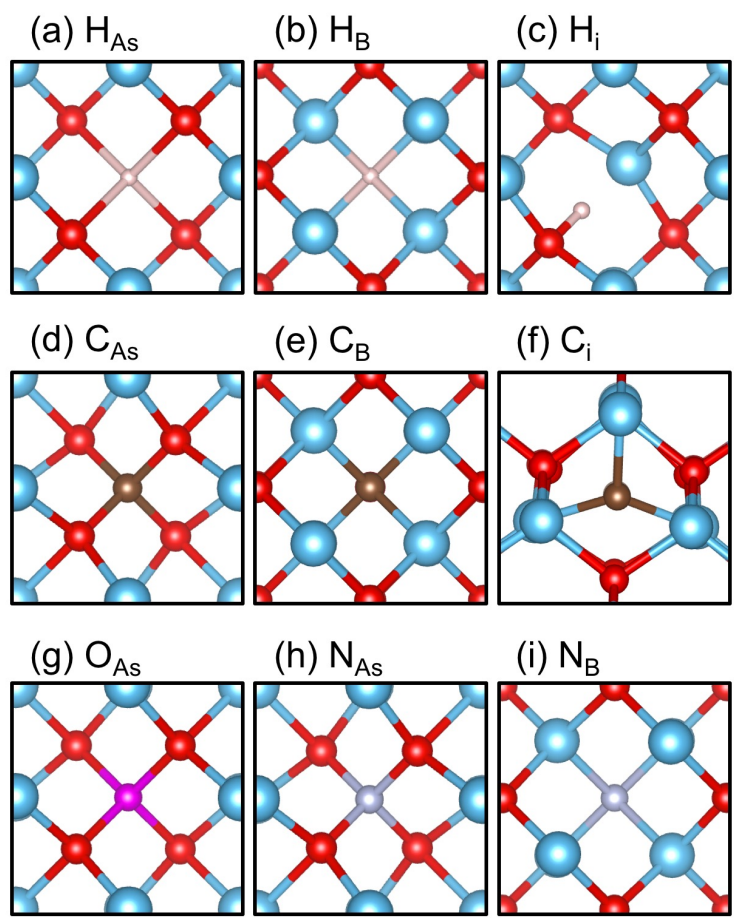

FIG. S2. Atomic structure of common impurities in BAs after structure optimization.

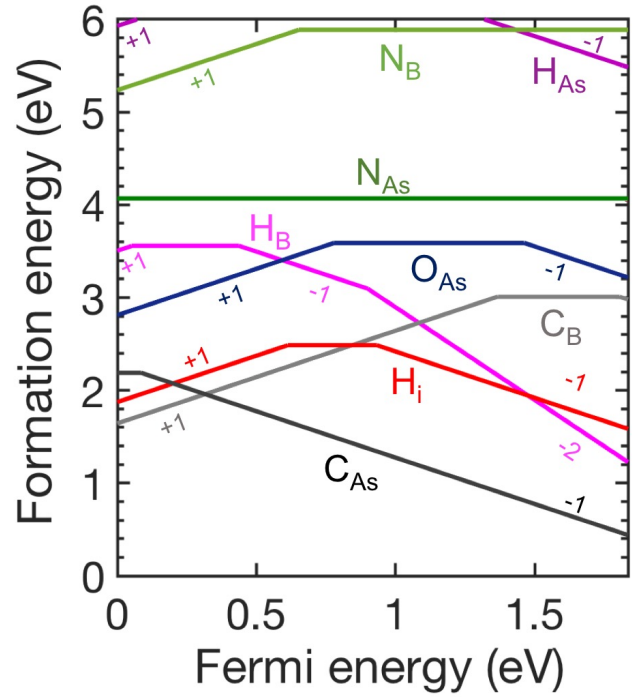

FIG. S3. Formation energy of common impurity defects as a function of the Fermi level. The chemical potential term for each defect is chosen to the value that yields the lowest formation energy for each defect.

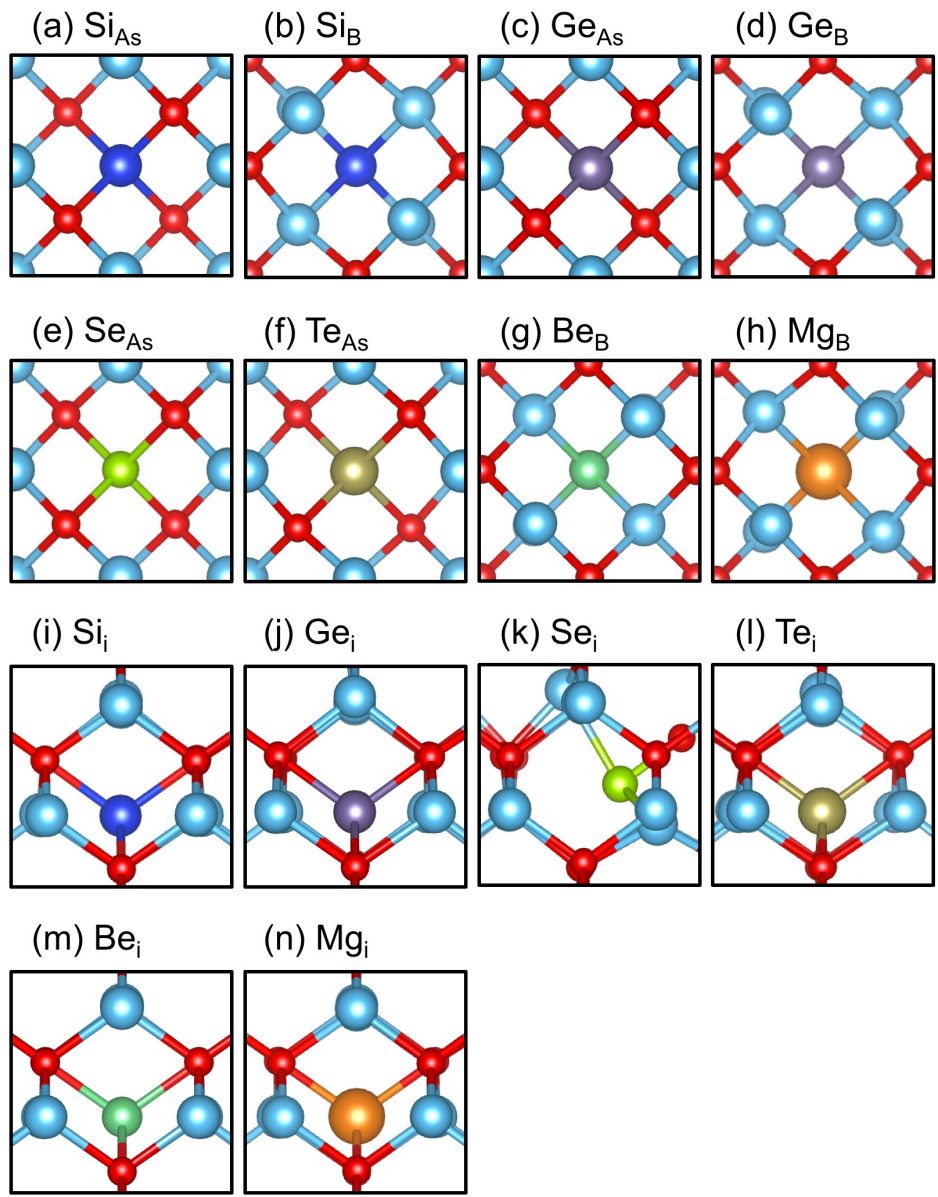

FIG. S4. Atomic structure of extrinsic dopants in substitutional and interstitial sites after structure optimization.

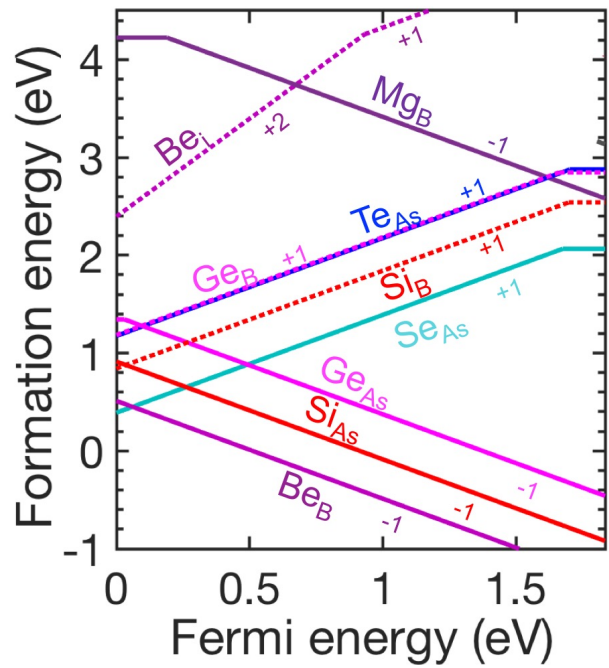

FIG. S5. Formation energy of donor and acceptor impurities as a function of the Fermi level. The chemical potential term for each dopant is chosen to the value that yields the lowest formation energy for each defect.

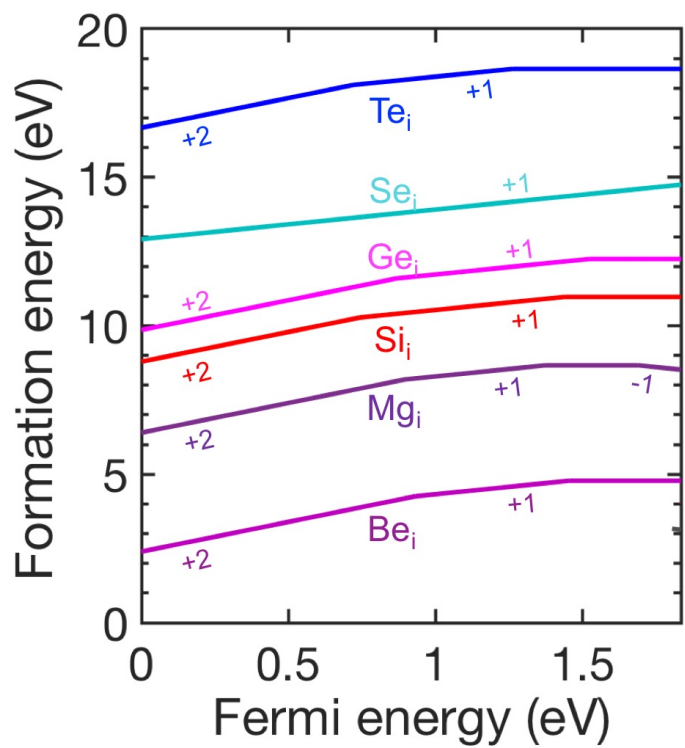

FIG. S6. Formation energy of extrinsic dopants incorporated into interstitial sites as a function of the Fermi level. The chemical potential term for each defect is chosen to the value that yields the lowest formation energy of each defect.